\documentclass[referee,sn-basic]{sn-jnl}% referee option is meant for double line spacing

\usepackage{graphicx}%
\usepackage{multirow}%
\usepackage{amsmath,amssymb,amsfonts}%
\usepackage{amsthm}%
\usepackage{mathrsfs}%
\usepackage[title]{appendix}%
\usepackage{xcolor}%
\usepackage{textcomp}%
\usepackage{manyfoot}%
\usepackage{booktabs}%
\usepackage{algorithm}%
\usepackage{algorithmicx}%
\usepackage{algpseudocode}%
\usepackage{listings}%
\newcommand{\tr}{\textcolor{black}}
\usepackage{multicol, multirow}

\theoremstyle{thmstyleone}%
\theoremstyle{thmstyletwo}%

\theoremstyle{thmstylethree}%

\begin{document}

\title[Polyspectral Clustering]{Polyspectral Mean based Time Series Clustering of Indian Stock Market}


\author[]{\fnm{Dhrubajyoti} \sur{Ghosh}}\email{dg302@duke.edu}



\affil[]{\orgdiv{Department of Biostatistics and Bioinformatics}, \orgname{Duke University}, \orgaddress{\street{242 Erwin Road}, \city{Durham}, \postcode{27705}, \state{NC}, \country{USA}}}

% \affil[2]{\orgdiv{Department}, \orgname{Organization}, \orgaddress{\street{Street}, \city{City}, \postcode{10587}, \state{State}, \country{Country}}}

% \affil[3]{\orgdiv{Department}, \orgname{Organization}, \orgaddress{\street{Street}, \city{City}, \postcode{610101}, \state{State}, \country{Country}}}


\abstract{In this study, we employ k-means clustering algorithm of polyspectral means to analyze 49 stocks in the Indian stock market. We have used spectral and bispectral information obtained from the data, by using spectral and bispectral means with different weight functions that will give us varying insights into the temporal patterns of the stocks. In particular, the higher order polyspectral means can provide significantly more information than what we can gather from power spectra, and can also unveil nonlinear trends in a time series. Through rigorous analysis, we identify five distinctive clusters, uncovering nuanced market structures. Notably, one cluster emerges as that of a conglomerate powerhouse, featuring ADANI, BIRLA, TATA, and unexpectedly, government-owned bank SBI. Another cluster spotlights the IT sector with WIPRO and TCS, while a third combines private banks, government entities, and RELIANCE. The final cluster comprises publicly traded companies with dispersed ownership. Such clustering of stocks sheds light on intricate financial relationships within the stock market, providing valuable insights for investors and analysts navigating the dynamic landscape of the Indian stock market.}

\keywords{Polyspectra; Bispectra; Clustering; Gap Statistic}



\maketitle

\section{Introduction}
\label{sec:intro}

The Indian stock market, a crucial component of the nation's financial landscape, has been a subject of great intrigue and scholarly exploration for decades. With its rich history dating back to the early 19th century, the Indian stock market has evolved into a complex and dynamic ecosystem that plays a pivotal role in the country's economic growth and development \citep{yadav2017stock, bala2013indian}. The importance of understanding the Indian stock market cannot be overstated. It serves as a barometer of economic health, reflecting the sentiments and expectations of investors, both domestic and international, while also serving as a source of capital for businesses to fuel growth and innovation. India's journey from a nascent market to one of the world's most dynamic and promising investment destinations is a testament to its resilience and adaptability in the face of global economic changes.

The clustering of stocks \citep{fama1992cross, nanda2010clustering} is a notable phenomenon in financial markets where stocks with similar characteristics or behaviors tend to group together, often exhibiting co-movements in their prices. This phenomenon has been extensively studied and observed in various stock markets worldwide. One of the most prominent forms of clustering in stock markets is sectoral clustering. Stocks within the same sector or industry often exhibit a tendency to move in concert due to shared economic drivers and market influences. For example, during economic downturns, stocks in cyclical sectors such as automotive or construction may experience simultaneous declines, while those in defensive sectors like healthcare or utilities may demonstrate relative stability. This sectoral clustering effect has been documented in studies across various stock markets \citep{sharma2017sectoral}. 
Another dimension of clustering is size-based clustering, where stocks are grouped based on market capitalization. Large-cap, mid-cap, and small-cap stocks tend to exhibit distinct characteristics and performance patterns. Research has shown that investors often create portfolios that cluster within these market capitalization categories, with each category attracting investors with different risk-return preferences  \citep{chen1991financial}. Risk is a critical factor in stock market clustering. Stocks with similar risk profiles, whether related to business risk, financial leverage, or market sensitivity, tend to cluster together. During periods of market turbulence or economic uncertainty, stocks perceived as riskier may exhibit stronger correlations in their price movements, reflecting investors' flight to safety. Geographical factors also play a role in clustering, especially in international stock markets. Stocks of companies based in the same region or country may share common exposure to local economic conditions, regulatory changes, or geopolitical events, leading to clustering based on geographical proximity \citep{bhutto2020portfolio}. Understanding these clusters of stocks is essential for investors and portfolio managers as it influences diversification strategies, risk management, and investment decision-making. By recognizing and analyzing the various dimensions of clustering, market participants can build more robust portfolios and navigate the complexities of financial markets more effectively.

The Indian stock market, characterized by its diversity in terms of sectors, industries, and market capitalization, presents a fascinating landscape for studying the clustering of stocks. Sectoral co-movements in Indian stock market has been studied by \cite{sharma2017sectoral}. Another dimension of clustering in the Indian stock market is market capitalization-based clustering. Stocks are commonly classified into large-cap, mid-cap, and small-cap categories based on their market capitalization. 
% Research by Chokshi and Pandey (2017) indicates that investors tend to cluster their investments within these categories. 
Large-cap stocks are considered less risky and more stable, attracting risk-averse investors, while mid-cap and small-cap stocks often entice risk-tolerant investors seeking higher returns. This leads to clusters of stocks within each market capitalization category \citep{singh2021empirical}.

\tr{The clustering technique mentioned beforehand primarily utilized features that capture first order features of time series. However when the time series is nonlinear, there is a plethora of information that is missed in the first order features like features obtained from spectral density or ARMA fit. Nonlinearity in stock market has long been a topic of research. \cite{hartman2018nonlinearity} explored nonlinearity in stock networks, while \cite{reboredo2012nonlinearity} discusses nonlinearity in forecasting returns. Nonlinearity in stock forecasting has also been analyzed in Indian stock market by \cite{siddiqui2015developing}. However, exploring the nonlinearity in time series clustering has been sparse. In order to do a nonlinear clustering, we need to consider features obtained from higher order polyspectra \citep{ghosh2024polyspectralmeanestimationgeneral}, which provided information on higher order interactions which are non-zero in a nonlinear time series. In this paper, we have extracted features from bispectra to provide a time series clustering of Indian Stock markets.}

Another way of clustering the stocks are based on time series clustering. Time series-based clustering of stocks is a powerful analytical approach that leverages historical price and trading volume data to identify patterns, relationships, and similarities among stocks over time. This methodology has gained significant attention in the field of finance and investment due to its potential to uncover valuable insights for investors, portfolio managers, and financial analysts. Time series-based clustering involves the use of statistical and machine learning techniques to analyze the historical price and volume data of stocks. The process typically involves extracting features from the time series and devise some machine learning techniques for clustering the time series. This temporal clustering of stocks can be used for portfolio construction (investors can build diversified portfolio by selecting stocks from different clusters \citep{ren2017dynamic}), risk assessment (stocks within the same cluster may exhibit comparable volatility patterns), Sector Rotation Strategies, behavioral analysis and many others. \cite{bollen2011twitter} utilized time series clustering to identify investor sentiment patterns in the U.S. stock market, revealing how sentiment-driven clusters of stocks can provide insights into market sentiment dynamics.

In this paper, we will propose a clustering approach based on spectral and bispectral means, using the higher order spectral features from a time series. The spectra of a time series are given by a Fourier transform of the autocumulants of a time series and carry significant information that might not be possible to obtain from the frequency-domain information. However, the spectra or the power series lacks information if the time series is non-linear or non-Gaussian, in which case we need to study higher-order autocumulants, which contain information that might be missing in power spectra for nonlinear time series. In-depth analysis of nonlinear time series using higher order cumulants have been done in \cite{mcelroy2023quadratic}. The Fourier transform of higher-order autocumulants is known as Polyspectra \citep{brillinger2011introduction}, and a statistic that can be obtained using a polyspectra is the polyspectral mean, which is a weighted average of a polyspectra. The asymptotic distribution and properties of polyspectral mean are explored in \cite{ghosh2024polyspectralmeanestimationgeneral}. Suppose the second order autocumulant is given by $\gamma(h_1, h_2) = E \{X_t X_{t+h_1} X_{t+h_2}\}$ (Second and third order autocumulants are same as second and third order moments, the two being different for order higher than $3$). Then the bispectra is given by:
\begin{align*}
    f_2(\lambda_1, \lambda_2) &= \sum_{h_1, h_2 \in \mathbb{Z}^k} \gamma(h_1, h_2)e^{-\iota \lambda_1 h_1 - \iota \lambda_2 h_2}
    % \int_{-\pi}^\pi \int_{-\pi}^\pi 
\end{align*}
The bispectral mean with weight function g is then given by:
\begin{align*}
    M_g(f_2) = \int_{[-\pi, \pi]^2} g(\lambda_1, \lambda_2)f_2(\lambda_1, \lambda_2)d\lambda_1 d \lambda_2
\end{align*}

In a parallel vein, the exploration of polyspectral means offers a promising avenue for gaining nuanced insights into the temporal characteristics of time series data within the spectral domain. To the best of the author's knowledge, the utilization of polyspectral means within the context of time series clustering remains largely uncharted territory, marking a significant departure from existing approaches. This study pioneers the application of spectral and bispectral means in the clustering of time series data, unveiling a novel dimension in the analysis of temporal patterns and spectral features.

\section{Data and Methods}

The data consists of the stock values of $49$ stocks collected from the NIFTY 50 stock market data. 
% We are considering the stocks that have data for the past $1000$ days, using which we will do our clustering analysis. 
\tr{The dataset is collected from \href{https://www.kaggle.com/datasets/rohanrao/nifty50-stock-market-data}{Kaggle}, where we have the price history and trading volumes of the fifty stocks in the index NIFTY 50 from NSE (National Stock Exchange) India. 
% All datasets are at a day-level with pricing and trading values split across .csv files for each stock along with a metadata file with some macro-information about the stocks itself. 
The data spans from 1st January, 2000 to 30th April, 2021. We have considered the last $1000$ days of the data for our clustering analysis, and we have used VWAP (Value-Weighted Average Price) for every stock.}

\tr{The dataset used in this study includes 49 out of 50 stocks from the NIFTY 50 index. The excluded stock is Bharti Infratel Limited, which underwent a significant structural transition following its merger with Indus Towers in 2020 \citep{vodafone2020merger}. As a result of this merger, Bharti Infratel ceased to exist as an independent publicly traded entity, leading to disruptions in its historical stock data availability. Due to this transition, the stock data for Bharti Infratel was unavailable in the dataset we used
% or inconsistent for the time period considered in this study
. Since our clustering methodology relies on continuous time series data, we opted to exclude this stock to maintain data integrity. The exclusion of a single stock does not materially affect the clustering results, as the remaining 49 stocks still provide a comprehensive representation of the NIFTY 50 market structure.
} 
\tr{In this study, we used a 1,000-day window as it strikes a balance between capturing long-term market structure while avoiding excessive short-term noise. This window length was chosen based on prior research in financial time series \citep{bustos2020stock, onnela2003dynamics, cao2012multifractal}, because sufficiently long horizons are required to extract stable spectral and bispectral features. Given that bispectral analysis inherently captures higher-order dependencies over time, we expect the clustering results to be relatively robust to moderate variations in the time window. While we did not explicitly test multiple window lengths (e.g., 500 or 1,500 days), this can be explored in future work to further validate the stability of the clustering structure.
}

% \tr{
\begin{table}[htbp]
    \centering
    \caption{Descriptive statistics of VWAP-scaled values for selected stocks. 
The table presents key time series characteristics, including \textbf{Mean Return} (average daily return), 
\textbf{Volatility} (measure of price fluctuations), \textbf{ACF Lag-1} (first-order autocorrelation capturing short-term dependencies), 
\textbf{Hurst Exponent} (indicating whether the stock follows a trend or mean reversion), 
\textbf{Trend Strength} (quantifying the dominance of long-term trends), 
and \textbf{Seasonal Strength} (measuring periodic patterns in stock price behavior).}
    \label{tab:descriptive_stats}
    \begin{tabular}{lccccccc}
        \hline
    Symbol & Mean Return & Volatility & ACF Lag-1 & Hurst Exp 
    % & Spectral Entropy 
    & Trend Strength 
    & Seasonal Strength 
    \\
    \hline
    ADANIPORTS & 0.004642 & 0.093273 & 0.124029 & 0.598035 
    % & 1.133761 
    & 0.913619 & 0.036247 \\
    ASIANPAINT & 0.003511 & 0.057868 & 0.083748 & 0.548264 
    % & 1.355831 
    & 0.969008 & 0.032016 \\
    AXISBANK & 0.001868 & 0.100149 & 0.150187 & 0.558980 
    % & 1.258382 
    & 0.981341 & 0.124283 \\
    BAJAJ-AUTO & 0.003011 & 0.129142 & 0.138089 & 0.566517 
    % & 1.135253 
    & 0.946395 & 0.150627 \\
    BAJAJFINSV & 0.003942 & 0.079524 & 0.216241 & 0.555013 
    % & 1.250950 
    & 0.961209 & 0.075671 \\
    BAJFINANCE & 0.003661 & 0.062111 & 0.198115 & 0.556876 
    % & 1.250544 
    & 0.973424 & 0.060793 \\
    BHARTIARTL & 0.002371 & 0.097661 & 0.145204 & 0.528287 
    % & 1.253974 
    & 0.978193 & 0.078211 \\
    BPCL & -0.003228 & 0.124760 & 0.066672 & 0.512610 
    % & 1.151792 
    & 0.930451 & 0.040754 \\
    BRITANNIA & 0.000074 & 0.101590 & 0.013219 & 0.548101 
    % & 1.379266 
    & 0.990516 & 0.084137 \\
    CIPLA & 0.002977 & 0.084454 & 0.138248 & 0.568832 
    % & 1.234805 
    & 0.954282 & 0.040107 \\
    COALINDIA & -0.002468 & 0.055536 & 0.181408 & 0.511234 
    % & 1.563858 
    & 0.986940 & 0.011804 \\
    DRREDDY & 0.002618 & 0.055181 & 0.200312 & 0.554359 
    % & 1.464788 
    & 0.979530 & 0.015755 \\
    EICHERMOT & -0.002500 & 0.076460 & 0.051470 & 0.478784 
    % & 1.422091 
    & 0.979651 & 0.020487 \\
    GAIL & -0.001877 & 0.054953 & 0.095367 & 0.530930 
    % & 1.885765 
    & 0.994930 & 0.017959 \\
    GRASIM & 0.001390 & 0.080409 & 0.109832 & 0.595154 
    % & 1.360918 
    & 0.975668 & 0.015299 \\
    HCLTECH & 0.000505 & 0.125211 & 0.027570 & 0.520015 
    % & 1.251389 
    & 0.980965 & 0.082445 \\
    HDFC & 0.003343 & 0.114447 & 0.156713 & 0.515941 
    % & 1.210587 
    & 0.956297 & 0.126470 \\
    HDFCBANK & -0.000063 & 0.092427 & -0.003916 & 0.542906 
    % & 1.453404 
    & 0.993651 & 0.043029 \\
    HEROMOTOCO & -0.000756 & 0.086942 & 0.160101 & 0.559905 
    % & 1.465072 
    & 0.994243 & 0.071202 \\
    HINDALCO & 0.003946 & 0.098586 & 0.107669 & 0.606707 
    % & 1.164991 
    & 0.936286 & 0.065586 \\
    HINDUNILVR & 0.003691 & 0.065341 & 0.050518 & 0.486033 
    % & 1.359173 
    & 0.966399 & 0.013248 \\
    ICICIBANK & 0.003394 & 0.084423 & 0.142444 & 0.537070 
    % & 1.234665 
    & 0.963242 & 0.139354 \\
    INDUSINDBK & -0.001097 & 0.057780 & 0.233733 & 0.584093 
    % & 1.759362 
    & 0.997403 & 0.016650 \\
    INFY & 0.001894 & 0.116513 & 0.020362 & 0.536065 
    % & 1.233794 
    & 0.970688 & 0.070211 \\
    IOC & -0.002817 & 0.061849 & 0.082809 & 0.504120 
    % & 1.494398 
    & 0.979694 & 0.016987 \\
    ITC & -0.001817 & 0.085460 & 0.149496 & 0.539463 
    % & 1.431885 
    & 0.987578 & 0.026277 \\
    JSWSTEEL & 0.006827 & 0.077552 & 0.181959 & 0.629366 
    % & 1.100032 
    & 0.862295 & 0.033625 \\
    KOTAKBANK & 0.003188 & 0.082927 & 0.133725 & 0.503933 
    % & 1.272980 
    & 0.971351 & 0.050764 \\
    LT & -0.001541 & 0.124670 & 0.083395 & 0.514035 
    % & 1.182158 
    & 0.961843 & 0.119631 \\
    M\&M & -0.001713 & 0.097023 & 0.009862 & 0.520331 
    % & 1.337180 
    & 0.981141 & 0.058622 \\
    MARUTI & 0.000282 & 0.099988 & 0.191167 & 0.577821 
    % & 1.247509 
    & 0.981835 & 0.097992 \\
    NESTLEIND & 0.002655 & 0.047111 & 0.084125 & 0.483298 
    % & 1.714718 
    & 0.986941 & 0.007416 \\
    NTPC & -0.001879 & 0.063117 & 0.091878 & 0.525328 
    % & 1.664169 
    & 0.992689 & 0.007106 \\
    ONGC & -0.001840 & 0.065145 & 0.114130 & 0.542353 
    % & 1.615488 
    & 0.992449 & 0.020283 \\
    POWERGRID & 0.001089 & 0.174913 & 0.026076 & 0.500386 
    % & 1.142787 
    & 0.947137 & 0.041866 \\
    RELIANCE & 0.001560 & 0.089587 & 0.093266 & 0.529852 
    % & 1.330346 
    & 0.984141 & 0.055302 \\
    SBIN & 0.001232 & 0.114421 & 0.212111 & 0.547330 
    % & 1.197035 
    & 0.971035 & 0.175840 \\
    SHREECEM & 0.003180 & 0.101310 & 0.153658 & 0.539488 
    % & 1.181244 
    & 0.949325 & 0.036920 \\
    SUNPHARMA & -0.000206 & 0.126199 & 0.153464 & 0.514365 
    % & 1.132222
    & 0.938217 & 0.033023 \\
    TATAMOTORS & -0.001306 & 0.048977 & 0.217036 & 0.582499 
    % & 1.666341 
    & 0.988146 & 0.018457 \\
    TATASTEEL & 0.004158 & 0.079340 & 0.180684 & 0.619734 
    % & 1.208656
    & 0.941291 & 0.074131 \\
    TCS & 0.001697 & 0.149814 & 0.027821 & 0.521904 
    % & 1.158895
    & 0.955090 & 0.067265 \\
    TECHM & 0.003397 & 0.071523 & 0.091635 & 0.548200 
    % & 1.328670 
    & 0.971540 & 0.057563 \\
    TITAN & 0.003774 & 0.068988 & 0.149385 & 0.534951 
    % & 1.289470 
    & 0.963365 & 0.037525 \\
    ULTRACEMCO & 0.003413 & 0.101803 & 0.134471 & 0.563086 
    % & 1.171401 
    & 0.944772 & 0.038998 \\
    UPL & -0.000915 & 0.096146 & 0.088367 & 0.557878 
    % & 1.332382
    & 0.989352 & 0.040765 \\
    VEDL & 0.000289 & 0.060452 & 0.129568 & 0.605508 
    % & 1.900387 
    & 0.999733 & 0.019261 \\
    WIPRO & 0.000103 & 0.137691 & 0.091157 & 0.524457 
    % & 1.108286 
    & 0.903840 & 0.092629 \\
    ZEEL & -0.001116 & 0.038295 & -0.000850 & 0.530639 
    % & 2.924925 
    & 0.995849 & 0.011984 \\
    \hline
    \end{tabular}
\end{table}
% }

We have only considered the Value-Weighted Average Price (VWAP) of every stock. VWAP is a key metric in the world of stock trading and finance. It represents the average price at which a particular stock has been traded throughout the trading day, weighted by the trading volume at each price level. VWAP is a critical indicator for both traders and institutional investors, helping them assess the efficiency of their trades and make informed decisions. By taking into account not only the stock price but also the trading volume, VWAP provides a more comprehensive picture of market trends and helps in identifying whether a particular trade was executed at a price favorable to the trader. VWAP is widely used in algorithmic trading, and it can also be used to assess the performance of portfolio managers. It is a valuable tool for optimizing trade execution and is instrumental in achieving best execution goals, particularly in the context of large and block orders \citep{almgren2005direct}. Additionally, VWAP can be applied to various timeframes, allowing traders to analyze short-term or intraday price trends as well as longer-term trends. By utilizing VWAP, market participants can better understand market dynamics and improve their trading strategies. \tr{Table \ref{tab:descriptive_stats} presents the descriptive statistics of VWAP-scaled values for selected stocks, capturing key time series characteristics. The Mean Return indicates the average daily return, while Volatility quantifies price fluctuations, reflecting market risk. ACF Lag-1 measures short-term dependencies in stock prices, with higher values suggesting stronger autocorrelation. The Hurst Exponent helps identify whether a stock follows a trend $(H > 0.5)$ or reverts to the mean $(H < 0.5)$. Trend Strength represents the dominance of long-term trends, and Seasonal Strength quantifies the presence of periodic patterns in stock price movements. These metrics provide insights into stock behavior, essential for clustering and further market analysis.}

We will utilize various information extracted from the time series. We will use spectral clustering techniques, where we will compute the spectral means and bispectral means for different weight functions, and use K-means clustering in order to provide an unsupervised clustering of the stocks. The spectral and bispectral means are computed according to the formula given in (\ref{eq:polyHat}), where the frequencies $\tilde{\lambda}$ are the Fourier frequencies, and $d(\lambda) = \sum_{t=1}^T X_t e^{-\iota \lambda t}$ is the Discrete Fourier Transform of the time series, and $\Phi(\lambda)$ is an indicator function that is $1$ iff the sum of no subset of $\lambda$ is divisible by $2\pi$, i.e. $\Phi(\underline{\lambda})$
 is one if and only if for all  subset $\lambda_{i_1}, \ldots, \lambda_{i_m}$ of $\underline{\lambda}$ ($1\leq m\leq k$),
 $\sum_{j=1}^m \lambda_{i_j} \not\equiv 0 \textrm{ (mod } 2 \pi)$.

\begin{align}
    \nonumber \widehat{M_g(f_k)} &\equiv  (2 \pi)^k T^{-k}\sum_{\tilde{\underline{\lambda}}} \hat{f}_k(\tilde{\underline{\lambda}})
    g\left(\tilde{\underline{\lambda}} \right)\Phi(\tilde{\underline{\lambda}}) \\
    &= (2 \pi)^k T^{-k-1} \sum_{\tilde{\underline{\lambda}}} d\left(\tilde{\lambda_1} \right)\cdots d\left(\tilde{\lambda_k} \right)d\left(-  [\tilde{\underline{\lambda}} ] \right)g\left(\tilde{\underline{\lambda}}\right)
    \Phi(\tilde{\underline{\lambda}}), 
    \label{eq:polyHat}
\end{align}

\begin{figure}[htbp]
    \centering
    \includegraphics[width = 0.6\textwidth]{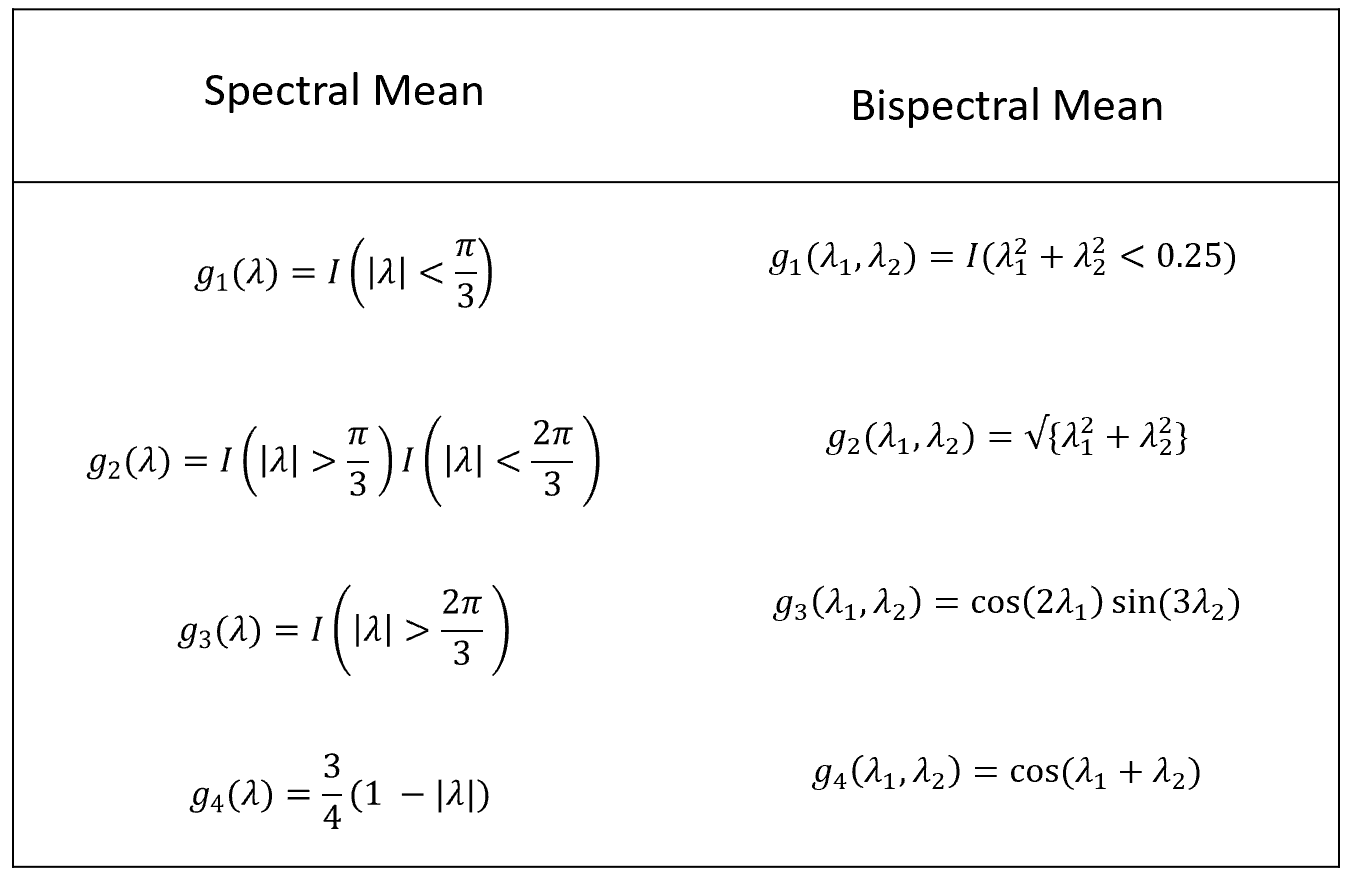}
    \caption{Weight functions taken for the spectral and bispectral means}
    \label{fig:weightFun}
\end{figure}

For example, when $k=1$, we are computing spectral mean, and for $k=2$. we are computing bispectral means. For this paper, we have only considered spectral and bispectral means, but higher order spectral means might also be used. The functions considered for the analysis are given in Figure \ref{fig:weightFun}. 

\tr{In this study, we employed a combination of spectral and bispectral weighting functions to enhance the stability and interpretability of our feature extraction process. For spectral estimation, we used indicator-based segmentation and a triangular window function to isolate different frequency bands while reducing spectral leakage. For bispectral estimation, we incorporated a circular indicator function to emphasize low-frequency interactions, a radial function to capture frequency magnitude relationships, and cosine-based functions to model nonlinear phase interactions.
. These weighting functions were selected to ensure that key higher-order interactions in the stock price series were preserved, minimizing distortion in the estimated polyspectral features. While alternative weighting schemes (e.g., Hanning or Gaussian windows) exist, our methodology primarily captures nonlinear dependencies, which are relatively stable across different choices of weighting functions. Future work could explore adaptive weighting strategies to further optimize feature extraction for time series clustering.
}

% The indicator functions to measure the spectral (and bispectral) content in the given region. The other functions measure several other spectral properties that can give us insights into the characteristics of the time series. 
We will consider the four spectral means, and four bispectral means (using the aforementioned weight functions) of the differenced time series. The differencing is done to gain stationarity since the theoretical asymptotic distribution of polyspectral means as derived in \cite{ghosh2022contribution} uses stationary time series. Additionally, we will also use some features like the period of the time series, Mean of the difference time series, Maximum of the difference time series, and the difference between the end and start of the difference time series. The obtained features are then used for unsupervised clustering using k-means algorithm.

\begin{figure}[h!]
    \centering
    \includegraphics[width=\linewidth]{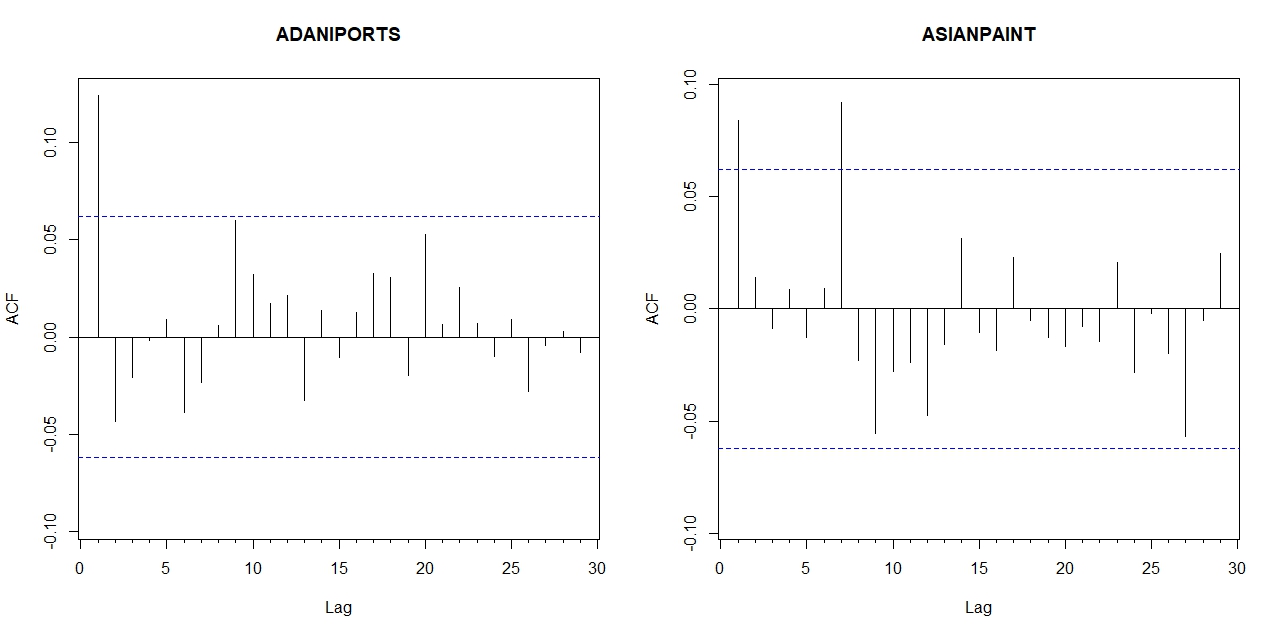}
    \caption{ACF Plot for two stocks -- ADANIPORTS and ASIANPAINT.}
    \label{fig:acf}
\end{figure}

\tr{To determine whether seasonality significantly affects stock clustering, we analyzed the autocorrelation function (ACF) plots of individual stock price series after differencing. Figure~\ref{fig:acf} shows the ACF plots for two representative stocks, ADANIPORTS and ASIANPAINTS. If strong seasonality were present, we would expect periodic spikes at fixed intervals (e.g., every 5, 22, or 252 lags corresponding to weekly, monthly, or yearly cycles in trading data). However, as seen in the plots, there are no consistent peaks at regular intervals, suggesting that strong seasonal effects are not present in these stocks. 
Since our clustering methodology relies on bispectral analysis, which inherently captures periodic structures and nonlinear dependencies, explicit seasonal adjustments were not necessary.}

\tr{Financial time series inherently contain market noise due to short-term fluctuations, high-frequency trading, and microstructure effects \cite{cont2001empirical}. While explicit noise filtering techniques, such as wavelet denoising or moving averages, were not applied, bispectral analysis inherently mitigates random noise by capturing higher-order dependencies rather than first-order fluctuations \cite{nikias1993higher}. To ensure that noise does not disproportionately influence clustering results, we used differencing to remove short-term trends and computed bispectral features over sufficiently long time windows (1,000 days), which naturally smooths out high-frequency fluctuations. Prior research has shown that bispectral estimates are robust to white noise components, as pure noise does not exhibit meaningful phase relationships in the bispectrum \cite{hinich1982testing}. Thus, while market noise is present in the dataset, its impact on clustering is minimal due to the properties of bispectral estimation and our preprocessing approach.
}

\section{Simulation}

Simulation studies of three set-ups are conducted to verify the validity of our method. We have generated time series using various procedures, and have applied k-means algorithm on the selected features.

\subsection{Simulation I (Binary-Class Stationary)}

\begin{figure}[htbp]
    \centering
    \includegraphics[width = 0.45\textwidth]{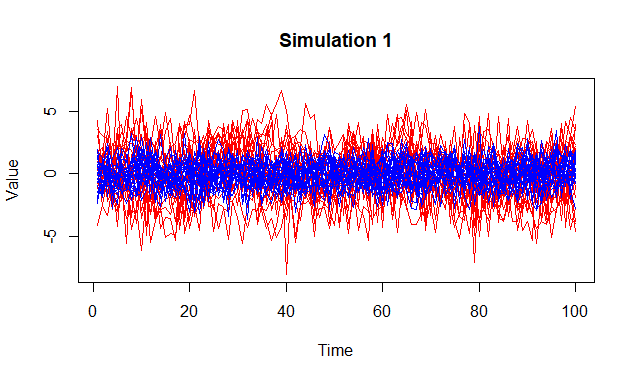}
    \includegraphics[width = 0.45\textwidth]{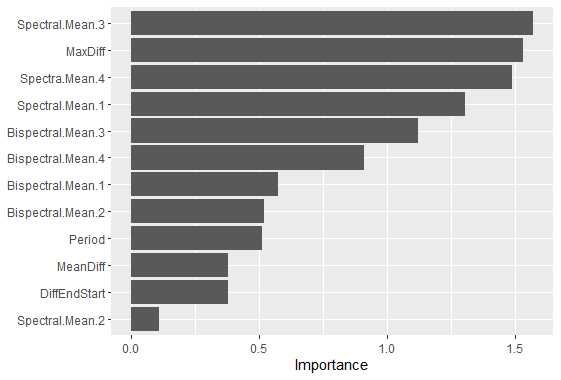}
    \caption{The left panel provides an instant of the group of 50 time series, and the right panel provided the corresponding feature importance.}
    \label{fig:sim1}
\end{figure}

The first simulation study has two groups of time series $Y_t^{(1)}$ and $Y_t^{(2)}$ with lengths $m$ and $n$ respectively. 
\begin{align}
    Y_t^{(1)} = X_t^{(1)} + \epsilon_t, t = 1, \ldots, T = 100,   Y_t^{(2)}  = X_t^{(2)} + \epsilon_t, t = 1, \ldots, T = 100
\end{align}
% \begin{itemize}
%     \item $Y_t^{(1)} = X_t^{(1)} + \epsilon_t$, $t = 1, \ldots, T = 100$,   $Y_t^{(2)}  = X_t^{(2)} + \epsilon_t$, $t = 1, \ldots, T = 100$
% \end{itemize}

where $X_t^{(1)}$ is generated from \textit{ARMA(2,2)} with AR coefficients $\phi = (0.1, 0.5)$ and MA coefficients $\theta = (0.2, 0.8)$. In other words:
\begin{align*}
    X_t^{(1)} &= 0.1 X_{t-1}^{(1)} + 0.5 X_{t-2}^{(2)} + W_t^{(1)} - 0.2 W_{t-1}^{(1)} - 0.8 W_{t-2}^{(1)}, \quad W_t^{(1)} \textrm{ is white noise.}
\end{align*}
$X_t^{(2)}$ is generated from \textit{GARCH(1,1)} with coefficients $0.2$ and $0.3$. In other words:

\begin{align*}
    X_t^{(2)} &= \sigma_t W_t^{(2)}, \quad W_t^{(2)} \textrm{ is white noise.} \\
    \sigma_t &= 0.2 X_{t-1}^{(2)2} + 0.3 \sigma_{t-1}
\end{align*}
The error terms $\epsilon_t$ are generated from $\mathcal{N}(0,1)$. The classification measures are given in Table \ref{tab:sim1_measure}, and one instance of the time series are given in Figure \ref{fig:sim1}.

\begin{table}[htbp]
    \centering
    \begin{tabular}{|cccccc|} \hline
        Sample Size  & Sensitivity & Specificity & F1-Score & Balanced Accuracy & AUC \\ \hline
         (30, 20) & $0.827$ & $1$ & $0.79$ & $0.931$ & 0.8846 \\ 
         (25,25) & $0.883$ & $0.967$ & $0.881$ & $0.923$ & $0.9227$ \\ 
         (10,40) & $0.813$ & $0.975$ & $0.921$ & $0.887$ & $0.9201$ \\
         (5, 45) & $0.815$ & $1$ & $0.889$ & $0.9$ & $0.9891$ \\
         \hline
    \end{tabular}
    \caption{Classification Accuracy Measures for 50 times series split into two groups of different sizes, each time series of length 100.}
    \label{tab:sim1_measure}
\end{table}

\subsection{Simulation II (Binary-Class with Different Trend)}

For the second simulation study, we will add a temporal trend in addition to the stationary noise to each of the groups.  The error terms are generated from $\mathcal{N}(0,1)$.

\begin{itemize}
    \item $Y_t^{(1)} = 10 \times \left( \frac{t}{T}\right)^2 + X_t^{(1)} + \epsilon_t$, $t = 1, \ldots, T = 100$
    \item $Y_t^{(2)} = 10 \times \left( \frac{t}{T}\right)^2 \times sin\left( \frac{0.9 \pi t}{T} \right) + X_t^{(2)} + \epsilon_t$, $t = 1, \ldots, T = 100$
\end{itemize}

\begin{figure}[htbp]
    \centering
    \includegraphics[width = 0.45 \textwidth]{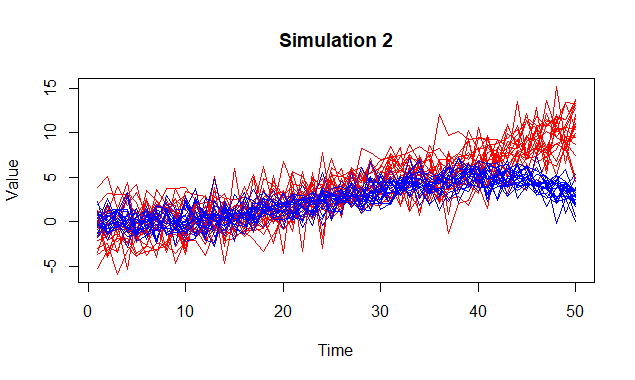}
    \includegraphics[width = 0.45 \textwidth]{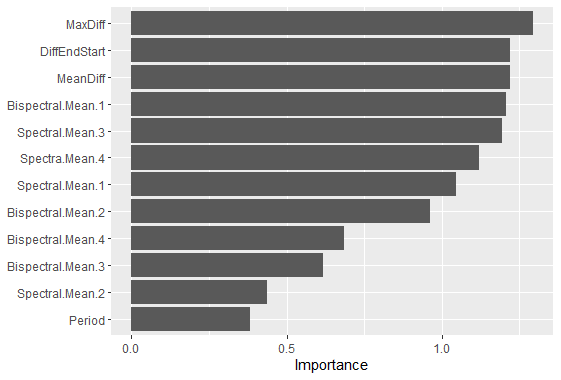}
    \caption{The left panel provides an instant of the group of 50 time series, and the right panel provided the corresponding feature importance.}
    \label{fig:sim2}
\end{figure}

In our last set-up, we saw that there is no seemingly distinguishing pattern of the time series, however, the classification would lie in the spectral domain, which was correctly captured by our spectral and bispectral means of different weight functions. In this set-up, the difference between the two groups is more visible, and for this set-up, we have obtained \underline{all accuracy measures to be $1$} for different group splits. An instance of the time series is given in Figure \ref{fig:sim2} along with a feature importance plot, which mentions the features most important for the clustering.

\subsection{Simulation III ($3$ groups with same or different trends)}

In this set-up, we will have time series from $3$ classes.

\begin{itemize}
    \item ($n_1 = 20$): $Y_t^{(1)} = 5 + X_t^{(1)} + \epsilon_t$, $t = 1, \ldots, T = 100$ 
    \item ($n_2 = 15$): $Y_t^{(2)} = 10 \times \left( \frac{t}{T}\right)^2 X_t^{(2)} + \epsilon_t$, $t = 1, \ldots, T = 100$
    % \item ($n_3 = 8$): $Y_t^{(3)} = 10 \times \left( \frac{t}{T}\right)^2 \times sin\left( \frac{0.9 \pi t}{T} \right)  X_t^{(3)} + \epsilon_t$, $t = 1, \ldots, T = 100$
    % \item ($n_4 = 7$): $Y_t^{(4)} = 10 \times \left( \frac{t}{T}\right)^2 X_t^{(4)} + \epsilon_t$, $t = 1, \ldots, T = 100$
    \item ($n_3 = 15$): $Y_t^{(3)} = 10 \times \left( \frac{t}{T}\right)^2 \times sin\left( \frac{0.9 \pi t}{T} \right)  X_t^{(3)} + \epsilon_t$, $t = 1, \ldots, T = 100$
\end{itemize}

Here $X_t^{(1)}$ and $X_t^{(2)}$ are generated using \textit{ARMA(2,2)} with coefficients same as the earlier simulations and $X_t^{(3)}$ is generated from \textit{GARCH(1,1)} with coefficients same as before. The error terms are generated from $\mathcal{N}(0,1)$.

\begin{figure}[htbp]
    \centering
    \includegraphics[width = 0.8\textwidth]{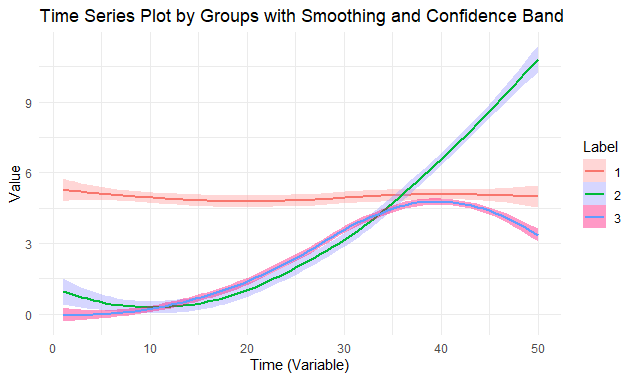}
    \caption{Time Series of 5 groups with different number of time series in each group}
    \label{fig:sim3}
\end{figure}

The time series plots of the $3$ groups are shown in Figure \ref{fig:sim3}. The accuracy measures are given below:

\begin{itemize}
    \item Sensitivity for the three classes: $(1, 0.87, 0.95)$
    \item Specificity for the three classes: $(0.89, 0.87, 0.93)$
    \item F-1 Score for each class: $(0.89, 0.90, 0.93)$
    \item Weighted F-1 Score: $0.945$
    \item Average AUC: $0.9541$
\end{itemize}

% \textbf{Discussion:} As we can see from our simulation studies, the k-means clustering using the features obtained provides good classification accuracy under different set-ups. In the first set-up, there was no difference in the mean trend, the only difference being in the underlying stationary distribution, with one coming from \textit{ARMA(2,2)}, a linear model, and the other coming from a nonlinear model, the \textit{GARCH(1,1)}. We have decent classification accuracy measures in this case, with almost all the group splits giving accuracy higher than $80\%$. In the second set-up, we have different mean trends along with different underlying stationary distributions, and here we have $100\%$ classification accuracy. Finally, in the last case, we have three groups, with different mean trends and different stationary distributions, and we again high classification accuracy close to $80\%$. All the simulations are done with a total number of $50$ time series, with each time series being of length $100$.

\begin{figure}[htbp]
    \centering
    \includegraphics[width = 0.6 \textwidth]{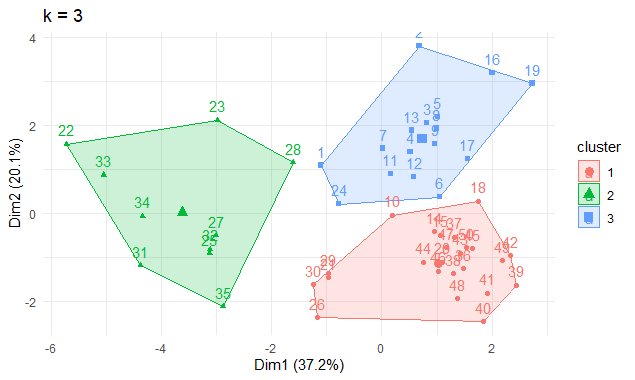}
    \caption{Results of k-means clustering of the three groups in Simulation Setting 3}
    \label{fig:3grp-cluster}
\end{figure}

\textbf{Discussion:} 
In the conducted simulation studies, the efficacy of k-means clustering was evaluated for its capacity to accurately classify time series data under a variety of conditions. The initial condition maintained homogeneity in mean trends across the series, with differentiation stemming solely from the inherent stationary distributions. Specifically, one distribution was modeled after an ARMA(2,2) process, emblematic of linear systems, while the counterpart was based on a GARCH(1,1) process, indicative of nonlinear dynamics. The classification accuracy under this condition was notably proficient, with the majority of classification measures surpassing $80\%$, with some measures going as high as $100\%$. The AUC was computed for different group splits and all of them came close to $90\%$.
Enhancing the complexity, the second simulation introduced heterogeneity in mean trends in addition to the distinct stationary distributions. This scenario yielded a classification accuracy of $100\%$, underscoring the k-means algorithm's sensitivity to divergent trend structures.
The complexity was further amplified in the final scenario by the introduction of three distinct groups, each characterized by their own mean trends and stationary distributions. The k-means clustering method continued to demonstrate high classification accuracy, with rates nearing $90\%$. The clustering result is given in Figure \ref{fig:3grp-cluster}.

These simulations were executed across 50 individual time series, each series comprising 100 data points. The findings underscore the robustness of the k-means clustering technique in discerning and classifying time series data, affirming its potential utility in statistical applications where underlying distributional and trend disparities are present.

\section{Data Analysis}

\begin{figure}[htbp]
    \centering
    \includegraphics[width = 0.45\textwidth]{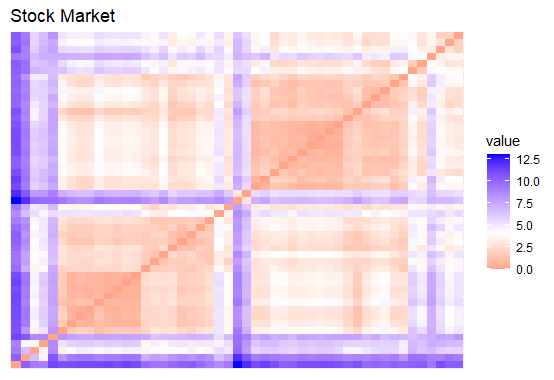}
    \includegraphics[width =0.45 \textwidth]{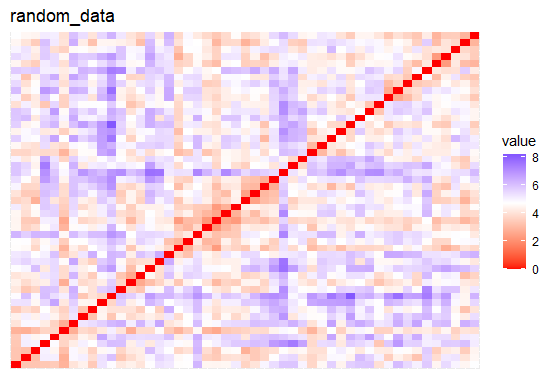}
    \caption{Dissimilarity matrix of stock market feature matrix (left panel) and a randomly generated dataset (right panel)}
    \label{fig:clustAbility}
\end{figure}

We will now apply our proposed algorithm to our dataset. However, before applying, we would try to first find out the possibility of presence of clusters in our dataset.  The clustering ability of a dataset is a crucial aspect of data analysis, particularly in unsupervised machine learning and data mining applications. It pertains to the inherent structure or grouping of data points within the dataset, which can be revealed through various clustering algorithms. One widely used metric for assessing the clustering tendency of a dataset is the Hopkins index, also known as the Hopkins statistic. The Hopkins index \citep{cross1982measurement} quantifies the degree to which a dataset can be partitioned into meaningful clusters by comparing the spatial distribution of actual data points to that of randomly generated data points. Specifically, the Hopkins index measures the probability that a given data point in the dataset is closer to another data point, typically one selected at random, than it is to the nearest neighbor in the dataset itself. A low value of the Hopkins index suggests a higher likelihood of meaningful clusters within the data, indicating that clustering algorithms may be effectively applied to uncover these patterns. The Hopkins index has gained popularity in the context of cluster analysis due to its simplicity and reliability in determining the feasibility of clustering operations \citep{kaufman2009finding}. The Hopkins index for our data is $0.7997$ with a p-value of $<0.001$, which indicates there is high possibility of clustering the dataset. We can also visualize the clustering tendency in Figure \ref{fig:clustAbility}, where we have plotted the dissimilarity matrix of our data (left panel) and the same for a randomly generated dataset (right panel). Here we can see there is some clear indication of the presence of patterns within the dataset.

The next step is to identify the correct number of clusters. Determining the optimal number of clusters in unsupervised classification is a critical aspect of ensuring meaningful and accurate clustering results. Various methods have been proposed in the literature to address this challenge, each offering unique insights into the structure of the data. The Elbow Method, introduced by 
% Ketchen and Shook 
\citep{ketchen1996application}, involves examining the rate of decrease in inertia or within-cluster sum of squares as the number of clusters increases. Silhouette analysis, as discussed by 
% Rousseeuw 
\cite{rousseeuw1987silhouettes}, evaluates the quality of clustering by measuring the separation and cohesion of clusters based on individual data point assignments. Gap Statistics, introduced by 
% Tibshirani, Walther, and Hastie
\cite{tibshirani2001estimating}, compare the performance of the clustering algorithm on the actual data to its performance on random data, helping identify an optimal cluster count. Other metrics such as the Davies-Bouldin Index \citep{davies1979cluster} and the Calinski-Harabasz Index \citep{calinski1974dendrite} assess the compactness and variance ratios of clusters, aiding in the determination of the correct number of clusters. It is recommended to utilize a combination of these methods, taking into account the specific characteristics of the dataset and the underlying clustering algorithm, to robustly identify the most suitable number of clusters for a given unsupervised classification task.

\begin{figure}
    \centering
    \includegraphics[width=0.8\textwidth]{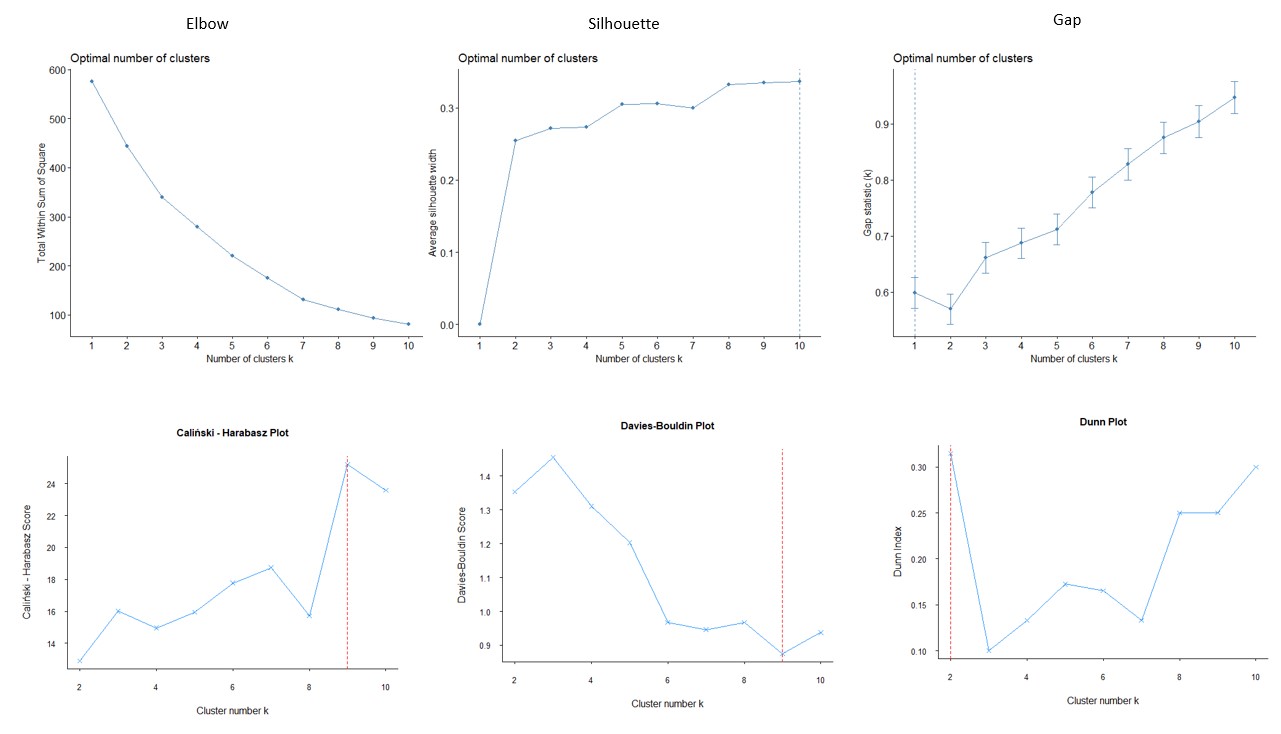}
    \caption{Different Measures of finding optimal number of clusters}
    \label{fig:optClust}
\end{figure}

\begin{figure}[htbp]
    \centering
    \includegraphics[width = 0.8\textwidth]{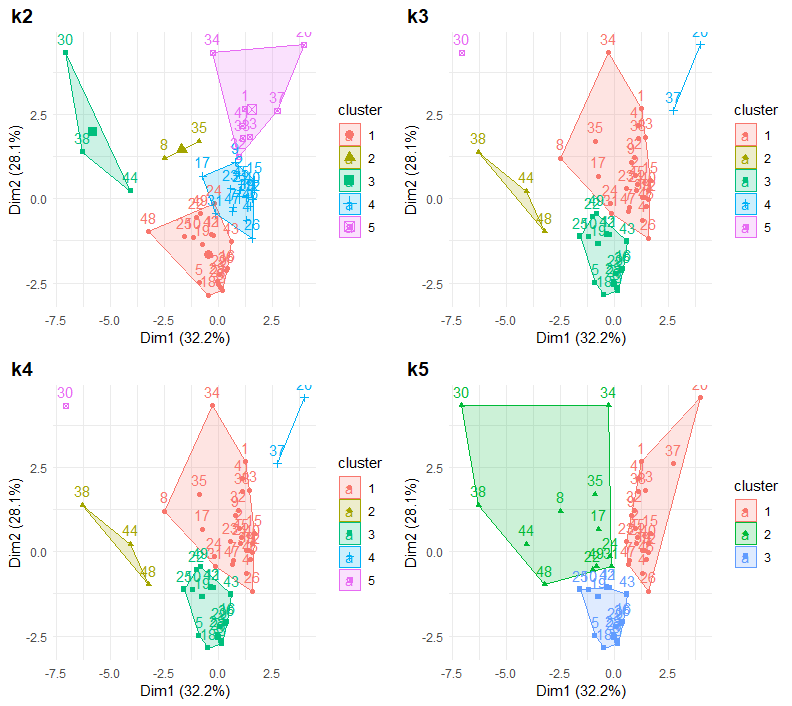}
    \includegraphics[width = 0.8\textwidth]{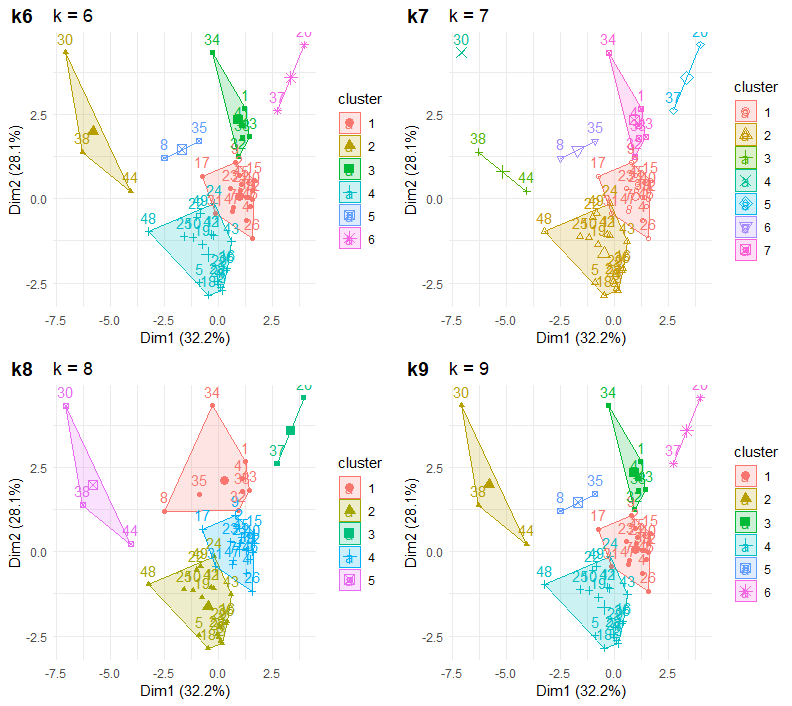}
    \caption{Clusters for different cluster sizes}
    \label{fig:cluster-all}
\end{figure}

Figure \ref{fig:optClust} provides the several metrics of clustering. As we can see, the elbow method doesn't provide any good estimate, and approximate estimate might be considered to be $3$. Silhouette index and the Gap Statistic is also continuously increasing showing that there is at least two clusters, but not able to find an optimal number of clusters. Tibshirani suggested in such cases to choose the cluster size $\hat{k}$ to be the smallest k such that $Gap(k) \geq Gap(k+1) - s_{k+1}$. Here $5$ seems to be a good number of cluster based on this rule.
% If we consider the rule to be the smallest $k$ such that $Gap(k) + s_{k} \geq Gap(k+1) - s_{k+1}$, then the optimal number of clusters would be $5$. 
The Dunn index proposes $2$ to be the optimal cluster size. Figure \ref{fig:cluster-all} provides the cluster for different choice of $k$ using the means algorithm.

As we can see from Figure \ref{fig:cluster-all}, clusters more than $5$ doesn't make much sense, while $2$ and $3$ seems to be too small number of clusters. Hence, we will now mainly focus on $5$ clusters. We will give information of clusters using two different algorithm, k-means, PAM (Partition Around Medoids), CLARA (Clustering Large Applications) and Fanny (Fuzzy Analysis CLustering). Partitioning Around Medoids (PAM) clustering, a variation of the k-medoids algorithm, represents a robust and interpretable approach to data clustering. Unlike traditional k-means clustering, PAM defines cluster centers as actual data points rather than the mean, making it less sensitive to outliers and noise in the dataset \citep{kaufman1990partitioning}. This medoid-based strategy ensures greater resilience to skewed distributions and enhances the algorithm's suitability for datasets with non-uniform cluster shapes. The algorithm iteratively refines clusters by selecting the most central point (medoid) within each cluster, optimizing the overall sum of dissimilarities between data points and their assigned medoids. PAM's focus on medoids, along with its flexibility in handling diverse data structures, renders it particularly valuable in scenarios where traditional centroid-based methods may falter, establishing its relevance across various applications in data clustering \citep{kaufman2009finding}.

Cluster Analysis of R (CLARA), an extension of the Partitioning Around Medoids (PAM) algorithm, offers an advantageous solution for handling large datasets by employing a sampling approach to representative subsets. CLARA, introduced by \cite{kaufman1990partitioning}, overcomes the computational challenges associated with PAM by selecting a subset of observations through repeated random sampling and applying PAM to each subset. This enables a more efficient approximation of medoids while maintaining the robustness of PAM's medoid-based clustering strategy. The resulting clusters are then aggregated to provide a comprehensive representation of the entire dataset. This unique combination of sampling and medoid-based clustering enhances CLARA's scalability, rendering it particularly suitable for larger datasets where the computational demands of exhaustive methods become prohibitive. 

Fuzzy Analysis by Non-Negative matrix factorization (FANNY) clustering stands out as a distinctive approach in the realm of cluster analysis by incorporating fuzzy logic principles. Introduced by \cite{kaufman1990partitioning}, FANNY extends traditional clustering methods by allowing data points to belong to multiple clusters simultaneously, reflecting the inherent uncertainty in many real-world datasets. By integrating non-negative matrix factorization, FANNY generates membership degrees for each observation across clusters, offering a nuanced representation of data ambiguity and facilitating a more flexible modeling of cluster assignments. The incorporation of fuzzy logic principles enhances FANNY's capacity to capture overlapping patterns in the data, making it particularly well-suited for scenarios where traditional hard clustering techniques may fall short in accommodating complex relationships among data points.

\begin{figure}[htbp]
    \centering
    \includegraphics[width = \textwidth]{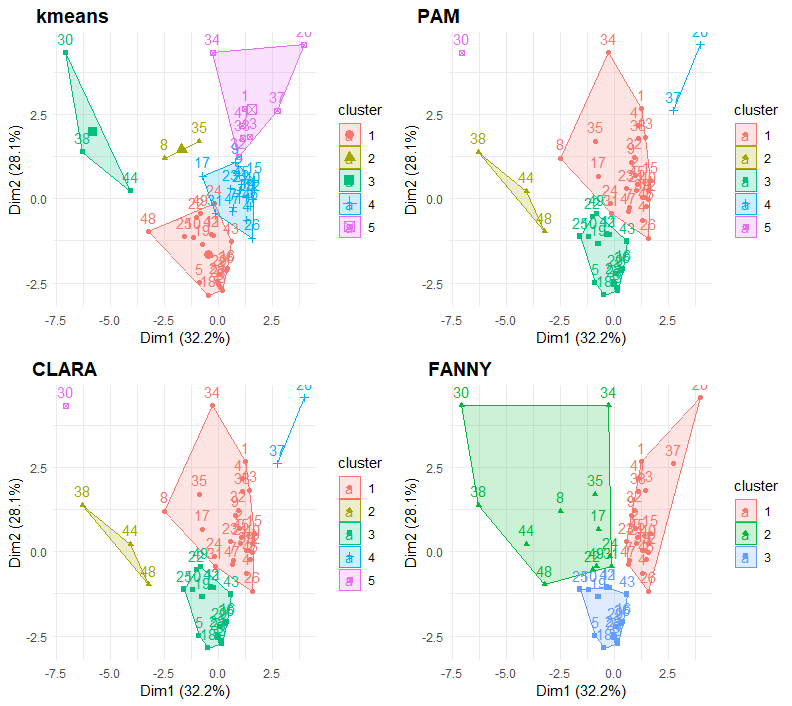}
    \caption{Clustering Results for the four types of clusters using number of desired clusters as 5}
    \label{fig:fourType}
\end{figure}

Figure \ref{fig:fourType} gives the clustering results of the four methods using the desired number of clusters to be $5$. It is to be noted here that the FANNY method gave three clusters even after giving the desired number to be $5$. Let us discuss mainly on the results of the k-means algorithm. One cluster contained WIPRO, TCS and POWERGRID, the first two being among the leading IT companies in India. The second cluster contains ADANIPORTS by ADANI group, TATASTEEL by TATA group, BAJAJ AUTO by BAJAJ Group, ULTRACEMCO and HINDALCO by BIRLA Group. Interesting observation was that government entity SBIN falls in the same cluster as this group, while other banks fall in the third cluster. The third cluster contains private banks like HDFCBANK, INDUSINDBK, and government owned companies like COALINDIA, BPCL, GAIL, NTPC, ONGC, etc. It seems these companies are mainly dealing with natural resources like coal, oil, etc. 
% The fourth group contains HCLTECH and SUNPHARMA, while the last group contains all publicly traded companies with ownership distributed among public shareholders, institutional investors, and promoters, like ASIANPAINT, AXISBANK, BAJAJFINSV, BAJFINANCE, BHARTIARTL, CIPLA, etc. Interestingly, RELIANCE also belonged to this group. 
\tr{The fourth group contains HCLTECH and SUNPHARMA, an unexpected pairing that suggests shared characteristics beyond sectoral classification, such as global market exposure, institutional investor preference, and nonlinear price behavior. The fifth group consists of large-cap, widely held companies spanning multiple industries, including ASIANPAINT, AXISBANK, BAJAJFINSV, BAJFINANCE, BHARTIARTL, and CIPLA, characterized by strong institutional ownership, high liquidity, and a mix of cyclical and defensive investment appeal. Notably, RELIANCE also appears in this group, reflecting its diverse business interests and market influence across multiple sectors.}
One can do further in-depth analysis of each of the clustering to unveil interesting trends. \tr{We have provided a brief discussion on the clusters in Section \ref{sec:disc}}.

% The full cluster information is given in Table \ref{tab:stock-info2}.

\begin{figure}[htbp]
    \centering
    \includegraphics[width = \textwidth]{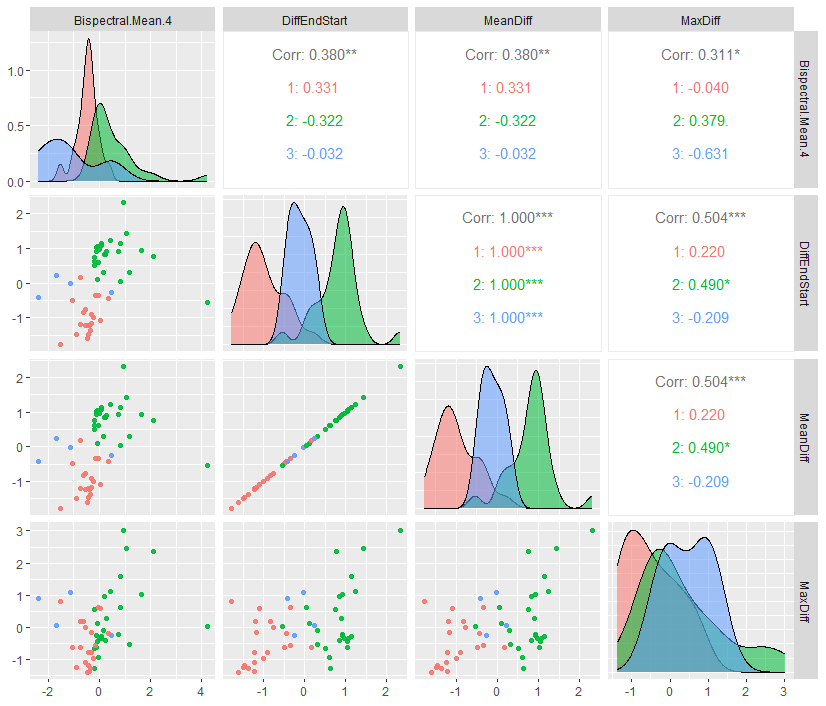}
    \caption{Featurewise Classification Power}
    \label{fig:featureClass3}
\end{figure}

Finally, we can see in Figure \ref{fig:featureClass3} the extent to which some of the features contributes to the classification, and the correlation between each features, along with the information whether they are significant or not. 
\tr{Figure \ref{fig:featureClass3} illustrates the feature-wise classification power of the selected clustering variables by displaying pairwise correlations, density distributions, and scatterplots of key statistical features derived from the VWAP-scaled stock price series. The diagonal elements show the distribution of each feature across clusters, while the off-diagonal elements provide insights into the relationships between different variables. One of the most critical observations is the distinct separation of clusters in terms of Bispectral Mean, indicating that higher-order spectral interactions significantly influence the clustering structure. Additionally, VWAP difference-based features—including DiffEndStart, MeanDiff, and MaxDiff—exhibit clear distributional distinctions across clusters, confirming that stocks with similar price movement variations are grouped together.
The correlation matrix in the upper triangle highlights notable relationships between features, with moderate to strong correlations between DiffEndStart, MeanDiff, and MaxDiff, suggesting that these metrics capture overlapping aspects of stock price dynamics. The positive correlation (0.380) between Bispectral Mean and DiffEndStart further indicates that nonlinear dependencies in stock price movements are linked to significant price changes over time. This provides strong evidence that the clustering algorithm is capturing fundamental price behavior characteristics rather than grouping stocks based on conventional industry or ownership structures. The scatterplots reveal that while some clusters exhibit clear separability in certain dimensions, others show overlapping patterns, reinforcing the idea that multiple interacting factors contribute to the clustering outcome.
Overall, Figure \ref{fig:featureClass3} validates that stocks are primarily grouped based on their statistical price behavior rather than traditional sector-based classifications. While some clusters align with business group affiliations, the results indicate that spectral properties, price fluctuations, and nonlinear dependencies play a dominant role in driving the clustering process. This analysis highlights the effectiveness of using frequency-domain and price-based statistical features to uncover latent patterns in stock market dynamics.}

% \begin{figure}[htbp]
%     \centering
%     \includegraphics[width = \textwidth]{Figures/br1.png}
%     \caption{Featurewise Classification Power}
%     \label{fig:featureClass1}
% \end{figure}

% \begin{figure}[htbp]
%     \centering
%     \includegraphics[width = \textwidth]{Figures/br3.png}
%     \label{fig:featureClass2}
% \end{figure}

\section{Discussion}
\label{sec:disc}

\begin{table}[h!]
% \small
\centering
    \begin{tabular}{|cccc|}
        \hline
        \textbf{Cluster} & \textbf{Stock} & \textbf{Major Stakeholders/Founders} & Serial Number \\
        \hline
        \multirow{8}{*}{1} & ADANIPORTS & Adani Group & 1\\
        & BAJAJ-AUTO & Bajaj family & 34\\
        & HINDALCO & Aditya Birla Group & 13 \\
        & JSWSTEEL & JSW Group & 20 \\
        & SBIN & Government of India & 32\\
        & SHREECEM & B.G. Bangur & 33\\
        & TATASTEEL & Tata Group & 37\\
        & ULTRACEMCO & Aditya Birla Group &41\\ \hline
        \multirow{19}{*}{2} & BPCL & Government of India & 48\\
        & BRITANNIA & Britannia Industries Limited & 49\\
        & COALINDIA & Government of India & 3\\
        & EICHERMOT & Eicher Motors Limited & 5\\
        & GAIL & Government of India & 6\\
        & HDFCBANK & HDFC Limited & 10 \\
        & HEROMOTOCO & Hero MotoCorp Limited & 11 \\
        & INDUSINDBK & Institutional and public shareholders & 16\\
        & IOC & Government of India & 18 \\
        & ITC & Institutional and public shareholders & 19 \\
        & LT & Larsen \& Toubro Limited & 22 \\
        & MARUTI & Suzuki Motor Corporation & 24 \\
        & M\&M & Mahindra Group & 25\\
        & ZEEL & Subhash Chandra & 27\\
        & NTPC & Government of India  & 28\\
        & ONGC & Government of India & 29\\
        & TATAMOTORS & Tata Group & 36\\
        & UPL & Institutional and public shareholders &  42\\
        & VEDL & Vedanta Limited & 43\\ \hline
        \multirow{3}{*}{3} & POWERGRID & Government of India & 30 \\
        & TCS & Tata Group &38\\
        & WIPRO & Azim Premji & 44\\ \hline
        \multirow{2}{*}{4} & HCLTECH & Shiv Nadar Foundation & 8 \\
        &SUNPHARMA & Dilip Shanghvi & 35 \\ \hline
%     \end{tabular}
%     \caption{Major Stakeholders/Founders of Selected Stocks I (Groups 1-4)}
%     \label{tab:stock-info1}
% \end{table}

% \begin{table}[htbp]
% % \small
% \centering
%     \begin{tabular}{|cccc|}
%         \hline
%         \textbf{Group} & \textbf{Stock} & \textbf{Major Stakeholders/Founders} & Serial Number \\
%         % \hline
%         \hline
        \multirow{19}{*}{5} & ASIANPAINT & Asian Paints Limited & 12\\
        & AXISBANK & Private shareholders & 23\\
        & BAJAJFINSV & Bajaj Finserv Limited & 45\\
        & BAJFINANCE & Bajaj Finance Limited & 46\\
        & BHARTIARTL & Bharti Enterprises & 47\\
        & CIPLA & Institutional and public shareholders & 2 \\
        & DRREDDY & Dr. Reddy's Laboratories Limited & 4 \\
        & GRASIM & Aditya Birla Group & 7\\
        & HDFC & Institutional and public shareholders &  9 \\
        & HINDUNILVR & Unilever PLC & 14 \\
        & ICICIBANK & Institutional and public shareholders & 15 \\
        & INFY & Institutional and public shareholders & 17\\
        & KOTAKBANK & Kotak Mahindra Group & 21 \\
        & M\&M & Mahindra Group & 25\\
        & MARUTI & Suzuki Motor Corporation & 24 \\
        & NESTLEIND & Nestle S.A. & 26 \\
        & RELIANCE & Mukesh Ambani and family & 31 \\
        & TECHM & Mahindra Group &39 \\
        & TITAN & Tata Group &40\\
        \hline
    \end{tabular}
    \caption{Major Stakeholders/Founders of Selected Stocks}
    \label{tab:stock-info2}
\end{table}

\tr{Table \ref{tab:stock-info2} exhibits the groupings of NIFTY50 stocks obtained by our clustering algorithm.  Cluster 1 comprises large industrial conglomerates and core economic sectors, including Hindalco (Aditya Birla Group), JSW Steel (JSW Group), Tata Steel (Tata Group), Shree Cement (B.G. Bangur), UltraTech Cement (Aditya Birla Group), Adani Ports (Adani Group), and State Bank of India (SBIN). These companies are highly sensitive to infrastructure development, commodity price cycles, and government policies on industrial expansion. The inclusion of SBIN, a public-sector bank, suggests a strong financial linkage between infrastructure financing and industrial growth. Many of these firms receive government contracts and institutional investments, leading to correlated stock movements that extend beyond sectoral classifications. Additionally, these conglomerates often engage in joint ventures and cross-sectoral investments, reinforcing their interconnected market behavior. The clustering algorithm captures these nonlinear dependencies, revealing how macroeconomic forces shape stock price dynamics in a way that traditional industry-based classifications might overlook.}

\tr{Cluster 2 consists of a diverse mix of companies spanning energy, banking, automobiles, FMCG, and infrastructure, including BPCL, Britannia, Coal India, Eicher Motors, GAIL, HDFC Bank, Hero MotoCorp, IndusInd Bank, IOC, ITC, Larsen \& Toubro, Maruti, M\&M, NTPC, ONGC, Tata Motors, UPL, Vedanta, and Zee Entertainment. Unlike Cluster 1, which is dominated by industrial conglomerates focused on metals, cement, and ports, Cluster 2 contains a blend of both private and public-sector companies, making it more exposed to policy-driven economic shifts, interest rate fluctuations, and global commodity price trends. The presence of government-owned companies (BPCL, ONGC, NTPC, IOC, GAIL, Coal India) highlights the impact of state regulations, energy pricing policies, and privatization efforts, whereas firms like HDFC Bank, ITC, and Maruti represent consumer-driven private-sector growth.
The mix of industries in this cluster suggests a broader economic sensitivity, where companies react to monetary policy changes, fuel price volatility, and consumer demand patterns, rather than being tied to a single industrial cycle. The banking and energy firms in this group are heavily influenced by global interest rates and oil market fluctuations, while companies like ITC and Britannia provide defensive stability during economic downturns. The inclusion of Larsen \& Toubro (L\&T), a major infrastructure player, and automobile manufacturers like Maruti and M\&M further supports the idea that this cluster is characterized by exposure to both long-term capital investments and short-term economic cycles. Overall, Cluster 2 reflects a highly dynamic segment of the economy, shaped by both government policy interventions and private-sector growth trends, setting it apart from the commodity-driven and infrastructure-focused stocks in Cluster 1.}

\tr{Cluster 3 consists of TCS, Wipro, and Power Grid Corporation of India, bringing together two major IT service providers and a government-backed utility firm. While at first glance, this grouping appears unconventional, it highlights a common characteristic of stable revenue generation, low volatility, and strong institutional investor participation. TCS and Wipro operate through long-term IT service contracts, while POWERGRID earns regulated revenue from electricity transmission, making all three companies resilient to short-term economic cycles. Additionally, these firms attract significant foreign and institutional investments, which may contribute to similar price movement patterns driven by capital flows rather than sector-specific trends. Unlike Cluster 1 (commodity-driven firms) or Cluster 2 (government-private policy-sensitive firms), Cluster 3 represents stable, defensive stocks that are less exposed to macroeconomic shocks and market volatility, reinforcing their unique positioning in the market.}

\tr{Cluster 4 consists of HCL Technologies (HCLTECH) and Sun Pharmaceuticals (SUNPHARMA)—two companies from completely different industries, IT and pharmaceuticals, respectively. At first glance, this grouping appears unexpected, but deeper analysis reveals several hidden commonalities that likely influenced their clustering.
Both HCLTECH and SUNPHARMA are recognized for global innovation, having been featured in Forbes’ list of the world’s most innovative companies. Additionally, both firms have a strong international revenue base, with HCLTECH deriving a large portion of its earnings from North America and Europe, while Sun Pharma generates over 50\% of its sales from the U.S. and other global markets. This international presence makes both companies highly sensitive to global economic conditions and currency fluctuations, which could explain their shared price movement patterns.
A key factor in their clustering is their investment profile—both companies attract strong institutional investor participation due to their consistent cash flows, high profit margins, and stable growth trends. Unlike more volatile sectors such as metals or banking, IT services and pharmaceuticals tend to be defensive in nature, meaning they perform relatively well even during economic downturns. This characteristic makes them appealing to long-term investors, mutual funds, and FIIs, which could drive similar capital flow patterns and market behavior.}

\tr{Cluster 5 consists of a diverse set of large-cap companies spanning multiple industries, including Asian Paints, Axis Bank, Bajaj Finserv, Bajaj Finance, Bharti Airtel, Cipla, Dr. Reddy’s, Grasim, HDFC, Hindustan Unilever, ICICI Bank, Infosys, Kotak Mahindra Bank, Mahindra \& Mahindra, Maruti, Nestlé India, Reliance Industries, Tech Mahindra, and Titan. Unlike the previous clusters, which exhibit sectoral coherence (e.g., industrial groups in Cluster 1, government-linked firms in Cluster 2), Cluster 5 is characterized by its mix of financial, consumer, technology, and industrial stocks, suggesting a grouping driven by market capitalization, institutional investment, and overall economic resilience rather than industry classification alone.
A key commonality among these stocks is that they are market leaders within their respective industries and widely held by institutional investors. Companies like Reliance, Infosys, and HDFC are among the largest publicly traded firms in India, making them highly liquid and actively traded. Additionally, this cluster contains multiple private-sector banks (ICICI, Axis, Kotak Mahindra, HDFC) and consumer-oriented companies (Asian Paints, Hindustan Unilever, Nestlé, Titan), reinforcing the idea that these stocks represent India’s economic backbone rather than a single industry focus.
Another distinguishing factor of this cluster is its balance of cyclical and defensive stocks. While companies like Mahindra \& Mahindra and Maruti are cyclical stocks that rise and fall with economic trends, firms such as Hindustan Unilever, Nestlé, and Cipla are defensive stocks that provide stability during market downturns. This mix suggests that the clustering algorithm captured nonlinear relationships in price movement and capital flows, grouping companies based on risk-adjusted return similarities rather than simple sector-based associations.
Overall, Cluster 5 represents a diversified portfolio of India’s largest, most resilient, and institutionally preferred stocks, bringing together leaders across banking, FMCG, IT, healthcare, and automobiles. Their shared characteristics—strong market capitalization, deep institutional ownership, and a mix of cyclical and defensive behaviors—likely explain their clustering, highlighting how higher-order spectral analysis captures investment-driven similarities that traditional sector classifications may overlook.}

\tr{While the clustering results reveal meaningful groupings based on shared market dynamics, institutional investment patterns, and macroeconomic sensitivities, these explanations may not be the only factors driving the observed clusters. The underlying structure of stock price movements is influenced by a complex interplay of investor sentiment, sectoral interdependencies, liquidity cycles, and global economic conditions, which may not be fully captured through the current analysis. Additionally, the use of higher-order polyspectral features suggests that nonlinear dependencies and hidden correlations could be shaping the clusters in ways that are not immediately intuitive from traditional financial metrics. Future research could explore alternative feature extraction techniques, incorporate higher-frequency trading data, and conduct robustness checks with different clustering algorithms to further validate the reasoning behind these stock groupings. Understanding the full set of factors that drive such clustering could provide deeper insights into market structure, risk assessment, and portfolio optimization.}

\section{Conclusion}
\label{sec:concl}

\tr{This study applied k-means and PAM clustering algorithms to analyze 49 Indian stocks, using spectral and bispectral features to capture nonlinear dependencies in stock price behavior. Five clusters were identified, revealing market-driven rather than purely sector-based groupings. Cluster 1 consists of large industrial conglomerates like Adani, Birla, and Tata Groups, alongside major steel and cement firms, reflecting their shared exposure to infrastructure spending and commodity cycles. The inclusion of State Bank of India (SBI) highlights its financial link to industrial expansion and public-private economic ties. Cluster 2, a mix of public and private enterprises, includes BPCL, ONGC, ITC, and Maruti, with stocks sensitive to government policies, interest rate changes, and consumer demand shifts. Cluster 3, featuring TCS, Wipro, and POWERGRID, suggests that these low-volatility, high-cash-flow companies share stable revenue models and strong institutional investor backing, making them more defensive than sector-bound.}

\tr{Cluster 4 links HCL Technologies and Sun Pharmaceuticals, likely due to their strong global presence, institutional ownership, and nonlinear market behavior rather than industry similarities. Cluster 5 includes Reliance, Infosys, HDFC, and Titan, representing high-liquidity, large-cap stocks balancing cyclical and defensive investment strategies. These results underscore how higher-order spectral analysis reveals hidden market interdependencies beyond traditional classifications. While the clusters align with certain financial characteristics, they may not fully explain all underlying patterns. Future research should explore alternative feature extraction techniques, different clustering models, and high-frequency data to refine these insights further, offering a new perspective on stock market structures, risk assessment, and portfolio diversification.}

\section*{Data Availability}

The data used in this study is the NIFTY stock market data and can be found at \href{https://github.com/djghosh1123/StockMarketData}{https://github.com/djghosh1123/StockMarketData}.
 % \bibliographystyle{rusnat} 
 % \bibliography{cas-refs}

\section*{Acknowledgment}

The author thanks the Editor, the Associate Editor and the referee for helpful comments that substantially improved the quality of the paper.

\section*{Conflict of Interest}

On behalf of all authors, the corresponding author states that there is no conflict of interest.

\section*{Funding Declaration}

This research received no specific grant from any funding agency in the public, commercial, or not-for-profit sectors.

% \section*{Ethics and Consent to Participate declarations}
\textbf{Ethics and Consent to Participate declarations:} Not applicable.

\textbf{Consent to Publish declaration:} Not applicable.

\bibliography{sn-bibliography}% common bib file

\end{document}